\documentclass[journal,draftcls,onecolumn,english]{IEEEtran}
\usepackage{amsmath,amssymb,amsfonts}
\usepackage{graphicx}
\usepackage{textcomp}
\usepackage{multirow}
\usepackage{caption}
\usepackage{subcaption}
\usepackage{hyperref} 
\usepackage{blindtext}
\usepackage{algorithm,algpseudocode}

\def\BibTeX{{\rm B\kern-.05em{\sc i\kern-.025em b}\kern-.08em
    T\kern-.1667em\lower.7ex\hbox{E}\kern-.125emX}}

\begin{document}

\title{Anti-Sexism Alert System: Identification of Sexist Comments on Social Media Using AI Techniques}

\author{\IEEEauthorblockN{Rebeca P. Díaz Redondo}\\
\IEEEauthorblockA{\textit{atlanTTic Research Center, University of Vigo} Spain  rebeca@det.uvigo.es \\}
\and
\IEEEauthorblockN{Ana Fernández Vilas}\\
\IEEEauthorblockA{\textit{atlanTTic Research Center, University of Vigo} Spain  avilas@det.uvigo.es  \\}
\and
\IEEEauthorblockN{Mateo Ramos Merino}\\
\IEEEauthorblockA{\textit{DATASALUS  S. Coop Galega, 36201 Vigo, Spain} mateo@datasalus.com   \\}
\and
\IEEEauthorblockN{Sonia Valladares}\\
\IEEEauthorblockA{\textit{DATASALUS  S. Coop Galega, 36201 Vigo, Spain} sonia@datasalus.com  \\}
\and
\IEEEauthorblockN{Soledad Torres Guijarro}\\
\IEEEauthorblockA{\textit{atlanTTic Research Center, University of Vigo} Spain   \\}
\and
\IEEEauthorblockN{Manar Mohamed Hafez}\\
\IEEEauthorblockA{\textit{Arab Academy for Science, Technology and Maritime Transport (AASTMT) Giza, Egypt} m.mohamed@aast.edu \\}
}

\maketitle

\abstract{Social relationships in the digital sphere are becoming more usual and frequent, and they constitute a very important aspect for all of us. {Violent interactions in this sphere are very frequent, and have serious effects on the victims}. Within this global scenario, there is one kind of digital violence that is becoming really worrying: sexism against women. Sexist comments that are publicly posted in social media (newspaper comments, social networks, etc.), usually obtain a lot of attention and become viral, with consequent damage to the persons involved. In this paper, we introduce an anti-sexism alert system, based on natural language processing (NLP) and artificial intelligence (AI), that analyzes any public post, and decides if it could be considered a sexist comment or not. Additionally, this system also works on analyzing all the public comments linked to any multimedia content (piece of news, video, tweet, etc.) and decides, using a color-based system similar to traffic lights, if there is sexism in the global set of posts. We have created a labeled data set in Spanish, since the majority of studies focus on English, to train our system, which offers a very good performance after the validation experiments.}

  \begin{IEEEkeywords}
artificial intelligence (AI); natural language processing (NLP); sexism; violence against women; social networks 
  \end{IEEEkeywords}

\section{Introduction}
Among the multiple forms of violence against women, digital sexism is currently of concern, since it is emerging, strongly aligned with the widespread adoption of new technologies in all social spheres. As~occurs with other types of violence, the~gender aspect is a determining factor, and women suffer this type of behavior to a greater extent, transferring to the digital sphere the sexist comments they suffer in other contexts. This new means of violence against women particularly affects (i) vulnerable groups, such as adolescents, as~reflected in different studies \cite{charpak2015quality,molanes2018deep}, and~(ii) people especially exposed to public opinion, such as journalists, politicians, artists, influencers, etc. The~digital field is, indeed, an~attractive scenario for perpetrators of violence, since there is a clear perception of invisibility and/or anonymity. This aggravates the consequences for the victims, because (i) they perceive there is no safe space, since the element used to exercise violence is also the one used by the victim to communicate with their environment, and (ii) the viral dissemination of comments and harassment amplify the damage~caused. 

Within this context, our research project seeks to use the same new technologies that enable this new gender-based violence, to fight against it. Our specific goal, is to use artificial intelligence (AI) techniques to automatically detect sexist comments on social networks, which would allow the identification of entities and/or users who are using inappropriate (or even violent) language towards women. Therefore, our proposal would impact the following two actors: (i) website holders, who would have an automatic mechanism to identify content that could be classified as sexist and discriminatory towards women, and (ii) public administrations and/or organizations, which would have guidelines to define more appropriate action guides and reinforce training providing education on gender and~equality.

From a technical point of view, we apply natural language processing (NLP \cite{molanes2018deep}) techniques, which combines AI techniques and linguistics, to allow computers to understand natural language and its structure. In~recent years, NLP techniques, such us lemmatization, steaming, tokenizing, stop words, regular expressions, or vectorizing \cite{bird2009natural,bird2006nltk}, have substantially evolved, and they have been successfully used in different areas. In~fact, one of the most widely known, is the so-called sentiment analysis, whose main goal is to infer whether a tweet, a~review, or a comment expressed on a social network, has a positive, negative, or even neutral connotation, in~addition to establishing a magnitude for this associated sentiment \cite{mabrouk2020deep}. Sentiment analysis has been used in social networks to analyze political content, product reviews, and applied, for~instance, in~recommendation~engines. 

Therefore, there is a good technical background that supports our approach: providing a mechanism able to automatically identify potential sexist comments on social media. We have coined our solution as \emph{anti-sexism alert} that, similarly to a traffic light, provides three colors according the level of alert: (i) red, if~it clearly identifies a sexist comment, (ii) green, if~it clearly identifies a non-sexist comment, and (iii) orange, if~it would need more information (such as context or the whole exchange of messages) to conclude if it is a sexist comment or not. Therefore, the main contributions of our study are the following:

\begin{itemize}
    \item Obtain a data set of public comments from social media relevant to our analysis. We focused on comments posted on YouTube, Twitter, and newspapers for the following reasons: (i) these posts are public, (ii) they are easily accessible, from~a technical perspective, and~(iii) these posts have gained influence in the general media and might go viral, which can provoke a public and persistent harm to the victim. We worked with comments in Spanish, since there is a lack of studies in this field in our language. We obtained a data set with 87,683 comments from different social media sources, and a labeled data set of 4389 comments, to train the~system.

    \item Analyze the different NLP techniques, in order to identify the most suitable ones and the procedure (methodology) to apply them to our approach. We studied keywords, syntactic constructions, and~other textual features (e.g., lemmazitation, steaming, tokenizing, stopwords, regular expressions, vectorizing, etc.). Finally, we decided to apply transfer learning, and use a pre-trained model obtained from the Hugging Face model repository: the Hate-speech-CNERG/dehatebert-mono-spanish, which was pre-trained to detect hate speech in Spanish \cite{aluru2020deep}, and~perform a fine-grained process to train it specifically for sexist comments. {As will be discussed below, this context of sexist comments presents an imbalance in the labeling of the different classes. One of the reasons why the dehatebert-mono-spanish model was chosen, is because it is already pre-trained in a scenario such as hate speech, which also has the characteristic of class imbalance associated with it, i.e.,~most of the comments analyzed do not present hate speech.}
    
    \item Define and train a suitable AI model to identify sexist comments and build our \emph{anti-sexism alert}. Our system works along two complementary lines: First, it is able to decide if a given comment (post) can be considered sexist. Second, it is able to analyze the whole set of public comments linked to a source (piece of news, video, tweet, etc.) and conclude if the whole set is sexist or not (using a color-based alert system similar to the one used in traffic lights).

\end{itemize}

The remainder of this paper is organized as follows: First, Section~\ref{sec:related} summarizes other approaches related to ours, as~well as providing an overview of NLP techniques applied to text analysis. Then, we introduce our proposal in Section~\ref{sec:methodology}, detailing the data set in Section~\ref{sec:dataset} and the AI model in Section~\ref{sec:model}. A~summary of the tests performed is explained in Section~\ref{sec:analysis}. We discuss the results in Section~\ref{sec:discussion} and, finally, we provide our conclusions and future lines in Section~\ref{sec:conclusions}.

\section{Related~Work}
\label{sec:related}

Cyberbullying \cite{rosa2019automatic} has increased in recent years with the growing use of new  technologies and, in~particular, with~the proliferation of social network sites (SNS) and the so-called Web 2.0, where anyone can contribute content through comments on forums, news, video channels, or by creating blogs, for~example. 

In recent years, research work in the field of the automatic identification of cyberbullying, has focused on the application of artificial intelligence (AI) techniques. These usually combine natural language processing (NLP) techniques with machine learning (ML) techniques \cite{mahesh2020machine}. NLP \cite{chowdhary2020natural} is understood as the field of study dedicated to the understanding of language in an automated way, through a machine (computer) and, therefore, it is very relevant in any text-based interaction between a computer and a person. This discipline encompasses, and combines the use of, AI techniques with knowledge of linguistics. Recently, these disciplines have evolved substantially and there are language processing techniques (e.g., lemma, steaming, tokenizing, stopwords, regular expressions, vectorizing, etc.) \cite{bird2009natural,loper2002nltk} that have been used in different domains. All of them require text analysis at different levels: morphological or lexical, syntactic, semantic, and pragmatic, which are especially useful for keyword~identification. 

Therefore, the~use of these techniques usually requires prior training procedures that are highly language-dependent and need to take into account: syntax, idioms, slang, and insults, among~other features. In~fact, areas such as computational linguistics \cite{srinivasa2018natural}, which deals with natural language modeling through rule inference and statistical analysis, in order to serve as a basis for software-based solutions, acquire special relevance. In~any case, the~detection of hate speech and cyberbullying is often more effective if multimodal sources of information, such as text and images, are combined, although~the exclusive use of text achieves very promising results and is inherently less costly. It should be noted that, if~the problem of automatically analyzing natural language text is already complex by nature, the~detection of harassment patterns on social media is even more complex, given that we are dealing with short text, without~clear structure, interspersed with emoticons and, very often, with~misspellings or abbreviated expressions of~words.

One of the sub-areas within natural language processing, that adopts a special relevance in the automatic identification of cyberbullying, is the so-called sentiment analysis \cite{liu2020sentiment} These techniques use different methodologies to be able to infer the sentiment conveyed by a text fragment or sentence: whether they convey positive or negative sentiment and, in~some cases, it is also possible to identify neutrality and even, with~poorer results, irony \cite{zhang2019irony}. Without~going into too many technical details, sentiment analysis in NLP essentially combines the use of so-called bag-of-words (or sets of terms) and n-grams (sequence of n elements in a specific order, in~this case, n words) \cite{chauhan2020comprehensive}. It is essential to know, therefore, the~syntactic structure of the language and, of~course, the~sequences of words that are usually used together and in a certain order \cite{zhao2019towards}. There are even some strategies that attempt to perform this sentiment analysis while the text is being written, as~in \cite{duric2012feature}. However, the~main problem they present, is that very large data sets are required for their correct training \cite{abbasi2010selecting}. Therefore, this type of technology is strongly dependent on the language in which the text is being~written.
 
As previously mentioned, the~increase in social interactions through the internet, has also led to an increase in cyberbullying, and~this type of attitude increased with the arrival of the COVID-19 pandemic and the confinement for long periods of time of a high percentage of the population, with access to these technologies. Specifically, studies have emerged that addressed this situation for specific areas of cyberbullying, such as cyberracism and digital machismo, and~others. \cite{mabrouk2020deep} show work aimed at specific online sectors, such as news, extracting a categorization of derogatory comments that can be used as a basis for further~work.

It was following the emergence of the Black Lives Matter movement in 2013, when the impact of cyberracism on the internet and, specifically, in~the field of social networks, began to be analyzed. Precisely, and~using the hashtag \#BlackLivesMatter, a~study was conducted in 2021 \cite{rani2022efficient} to infer the emotions that were transmitted with comments on social networks, using two regions in the US (Minnesota and Washington D.C.) as a field of study, and achieving a result with very high accuracy (94\%) in the detection of feelings. Another very recent study, \cite{balakrishnan2022unravelling}, published in 2022, deals with the automatic identification of racist attitudes using machine learning methods from public interactions on the Twitter network. For~this purpose, material collected during the month of March 2020 (at the height of the pandemic) was used as the training material, and sentiment analysis mechanisms were applied \cite{chauhan2020comprehensive}.

Of equal concern in the research field, is the growing incidence of cyber-hostility directed towards women, simply because they are women. This concern is reflected, for~example, in~the growing number of publications related to this problem in the last two years in international journals devoted to gender issues, such as Gender \& Society ({ISSN} 
 0891-2432, \url{https://journals.sagepub.com/home/gas} (accessed on 20 January 2023)), Journal of Gender Studies ({ISSN} 
) 0958-9236, \url{http://miar.ub.edu/issn/0958-9236} (accessed on 20 January 2023), but~also in medical journals such as
Social Science \& Medicine ({ISSN} 
 0277-9536, \url{https://www.journals.elsevier.com/social-science-and-medicine} (accessed on 20 January 2023)), Women's Health Issues ({ISSN} 
 1049-3867, \url{https://www.journals.elsevier.com/womens-health-issues} (accessed on 20 January 2023)), and Women's Studies International Forum ({ISSN} 
 0277-5395 \url{https://www.journals.elsevier.com/womens-studies-international-forum} (accessed on 20 January 2023)).

These publications include studies on a wide range of topics: from social aspects and incidence in the family environment, to~aspects related to the impact on health (mental and physical, in~addition to the processes of somatization of emotional distress). Meanwhile, there has also been an increase in publications related to new technologies and natural language processing. For~example, in~its 2020 edition, the~international conference LREC (Language Resources and Evaluation Conference) dedicated a specific space called Trolling, Hostility and Cyberbullying, with a special mention to gender hostility on networks \cite{kumar2020evaluating}, where practical workshops on automatic detection of digital machismo were held in three languages: English, Bengali, and~Hindi. 

In fact, English is the language used in the majority of studies focused on identifying sexist comments. For~instance, in~\cite{sai2022explorative} the authors apply fusion techniques to detect multimodal hate speech, which constitutes a good starting point for other practices. In~another example \cite{salminen2018anatomy}, analyses are performed using Twitter data, to detect sexist and racist harassment in this~language.

In the field of artificial intelligence and natural language processing techniques applied to automatic detection of cyberbullying, there are some initiatives that address the problem in a general way. For~example, some pre-trained models to detect ”hatespeech” (or hate speech) in Spanish are available, such as Hate-speech-CNERG/dehatebert-mono-spanish \cite{cruz2022selecting}.

Fewer are those that attempt to perform this automatic detection in the specific field of digital machismo. In~any case, the~existing ones have achieved incipient and promising results, but~in other languages (mainly English). However, to~the best of our knowledge, there are no relevant initiatives in~Spanish.

\section{Methodology}
\label{sec:methodology}

\begin{figure}[htpb]
\centering
\includegraphics[scale = 0.80]{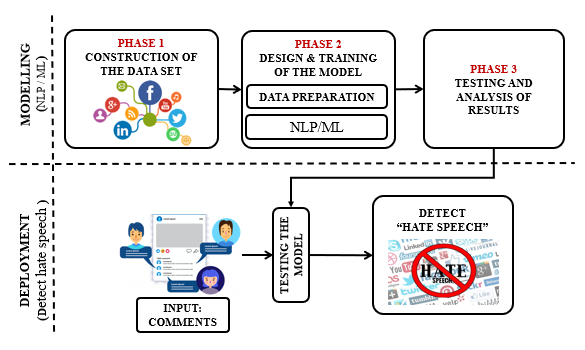}
\caption{Methodology of the \emph{anti-sexism alert} system.}
\label{fig:methodology}
\end{figure}

In order to obtain an \emph{anti-sexism alert} system, which should be able to classify text from social media, deciding whether or not it contains sexist comments, we propose a methodology (depicted in Figure~\ref{fig:methodology}) composed of three phases:

\begin{itemize}
   \item Phase 1: Construction of the data set. We firstly selected and gathered relevant data for our analysis, which was used to train and validate our proposal. We used scraping techniques to gather comments from different types of social media posts (newspaper, social media networks, video platforms, etc.). Part of the data was labeled, to train the system, as~detailed in Section \ref{sec:dataset}
   \item Phase 2: Design and training of the model. After~gathering and building the data set, we selected the most appropriate NLP techniques and defined how to apply them, by identifying relevant textual characteristics with gender perspective. Section \ref{sec:model} includes all the technical details of this~phase.
   \item Phase 3: Tests and analysis of the results. Finally, we tested our approach, to assess its performance as an indicator of sexist comments. The~obtained results are presented in Section \ref{sec:analysis}.
\end{itemize}

\section{Phase 1: Building the Data~Set}
\label{sec:dataset}

In order to obtain an adequate data set, we worked on two aspects. First, gathering comments that were publicly posted on different social media sources and, second, selecting a subset of comments to be manually labeled, in order to detect sexist comments (needed to train the final system).

\subsection{Gathering Data from Social~Media}
\label{sec:gathering}

After a detailed analysis, we decided to use three kind of social media sources. First, a~social media network (SNS). We selected {Twitter} 
 (\url{https://twitter.com/} (accessed on 20 January 2023)), since it is a public SNS with a huge impact worldwide, and~also in Spain. Second, a~video platform. We selected {YouTube} 
 (\url{https://www.youtube.com/} (accessed on 20 January 2023)), another website with an enormous impact. Finally, digital media (newspapers). We selected two Spanish newspapers: {"El Mundo"} 
 (\url{https://www.elmundo.es/} (accessed on 20 January 2023)), a~general mainstream digital newspaper, and~{"Marca"} 
 (\url{https://www.marca.com/} (accessed on 20 January 2023)), a~sports digital newspaper. The~former is among the top five most read generalist newspapers online in Spain, with a weekly readership of 13\%, and the latter is the most read sports newspaper in its digital version (weekly readership of 11\%) \cite{Medios22}.

In order to obtain comments from Twitter posts, comments on YouTube videos, and news comments in the newspapers, we selected content sources (posts/hashtags, videos, and pieces of news), taking into account the following aspects:

\begin{itemize}
    \item [-] Gender of the protagonist of the content source ("male" or "female"). For~instance, a~``female'' content source might be a piece of news about International Women's Day or the new video of a famous female~singer.

    \item [-] Number of protagonists (``individual'', ``collective'', ``hybrid''). The~first category (``individual'') refers to a content source about a specific person: for instance, a~hashtag about a politician. The~second category (``collective'') refers to a content source related to a generic topic: for instance, a~piece of news about the football league. Finally, the~third category (``hybrid'') refers to a content source that copes with both: individual and collective aspects, for~instance, a~YouTube video about the new Nobel Prize in literature, that also discusses the role of women in~literature.

    \item [-] Context of the content source (``professional'', ``personal'', ``hybrid''). The~first category (``professional'') includes those content sources related to a profession, for~instance, a~piece of news about the career of a famous tennis player or a YouTube video about the global soccer awards. The~second one (``personal'') includes those content sources related to private life: a famous woman is pregnant or not, for~instance. Finally, the~third category (``hybrid'') includes content sources that covers both aspects: a piece of news talking about the maternity leave of a Prime Minister and also how she faces work challenges.
\end{itemize}

We used two criteria to select content sources for our data set. First, trying to balance the content sources according to the previous characteristics: (i) in terms of gender of the protagonist (approximately half per type, in accordance with the number of comments), and (ii) in terms of number of protagonists and context (approximately one third within each type, in accordance with the number of comments). Second, we selected sources that had generated different number of comments (low and high), but~fixing a lower threshold of 15 (specially needed for newspapers, where the number of comments is low in general) and an upper threshold of 5000 (needed for Twitter and YouTube, where the posts are more frequent).

\begin{figure}[htpb]
\includegraphics[width=\textwidth]{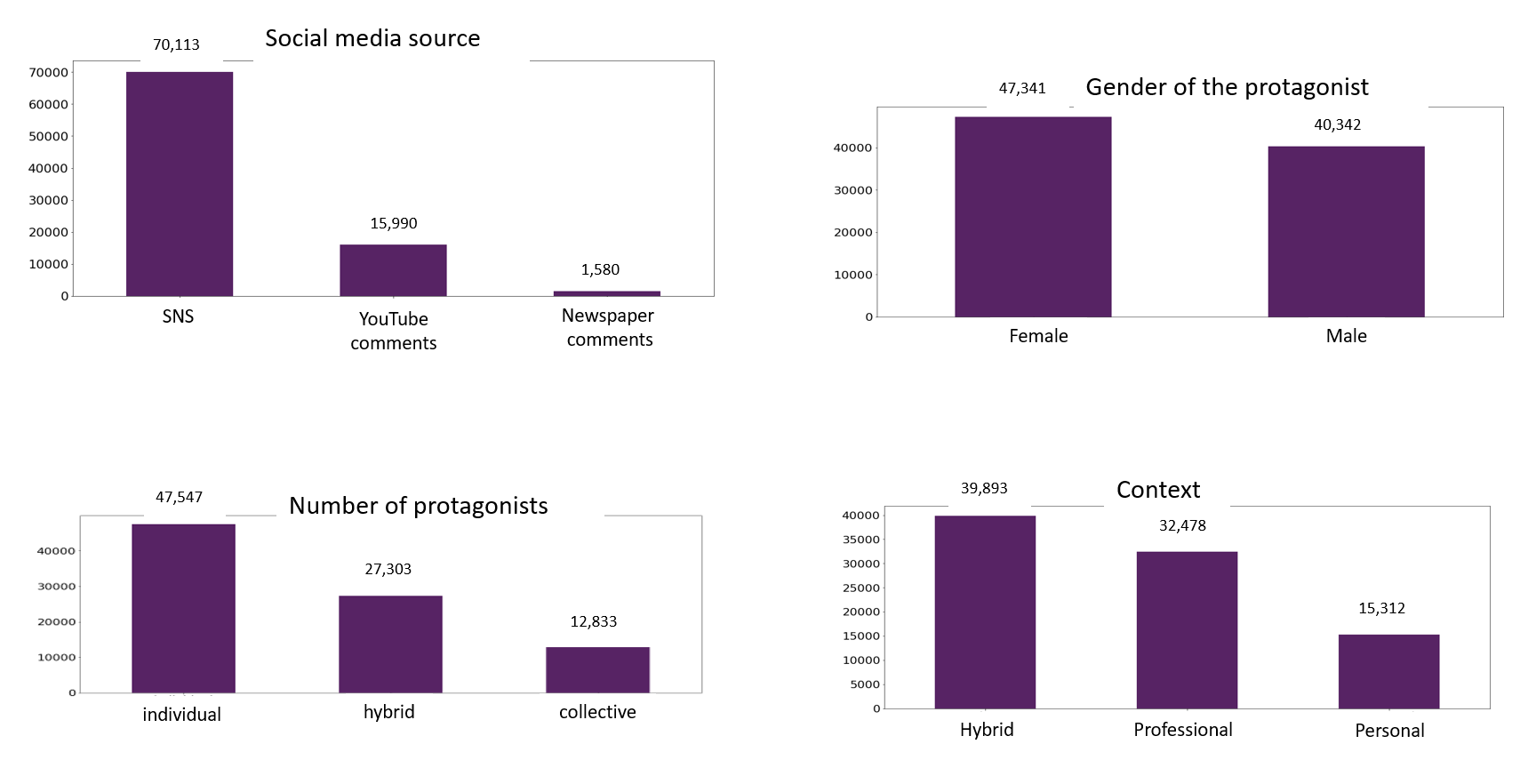}
\caption{{Characteristics} of the comments in the data~set.}
\label{fig:statistics1}
\end{figure}

To sum up, we selected $59$ content sources (Table \ref{tab:listSources} in Appendix  and gathered the comments linked to them, obtaining a total of $87,683$ comments, which are characterized in Figure~\ref{fig:statistics1}. We can observe the lowest number of comments come from newspapers and they are balanced according to gender. However, the~number of comments regarding individual persons is much higher than collective and hybrid. Besides, the~number of comments referring to the hybrid context (professional and personal) is higher than only professional or only personal. Therefore, and~in order to avoid bias when training the model, we corrected the unbalanced data in the labeling phase (Section \ref{sec:labelling}).

\subsection{Labeling Part of the Data Set for~Training}
\label{sec:labelling}

After having a complete Spanish data set from different media and characteristics, we selected a subset to create an appropriate training data set, which must be labeled according to the presence of sexism. This was essential to train our supervised learning system to eventually be able to automatically identify sexism in new~comments. 

In order to avoid an unbalanced training data set, we selected comments from approximately one third of each type of social media source (posts/hashtags, videos, and pieces of news). Besides that, we selected comments that assured a balanced representation of the gender of the protagonist (half of each), number of protagonists (one third of each), and context (one third of each). At~the end, we obtained 4389 balanced comments, approximately 5\% of the global data~set.

We need to highlight that, for this proof of concept, the~\emph{anti-sexism alert} system does not have contextual information of the comment to be classified. Therefore, we proceeded with the manual labeling process according to four~categories:
\begin{itemize}
    \item [-] Yes (sexist comment): clearly the comment has a sexist connotation.
    \item [-] No (sexist comment): clearly the comment does not have a sexist connotation.
    \item [-] Discard: the comment cannot be used for the analysis because of the following reasons: comments with a large number of spelling or grammatical errors; comments with a high presence of non-alphanumeric characters; comments without a clear intention or with a doubtful presence of sarcasm; comments in which we do not fully understand what the person wanted to say; duplicate comments; comments in a language other than Spanish; incomplete comments; and comments that are associated with other sources of information, such as images.
    \item [-] Depends on the context: it is not possible to decide if the comment has a sexist connotation or not without knowing the context or conversation.
\end{itemize}

The labeling process was performed by four different persons, of both genders. A~comment was labeled as ``sexist'' only if the majority agreed.  The few times that there was a tie, we decided to label the comment as ``depends on the context''.

\begin{figure}[htpb]
\centering 
\includegraphics[width=\textwidth]{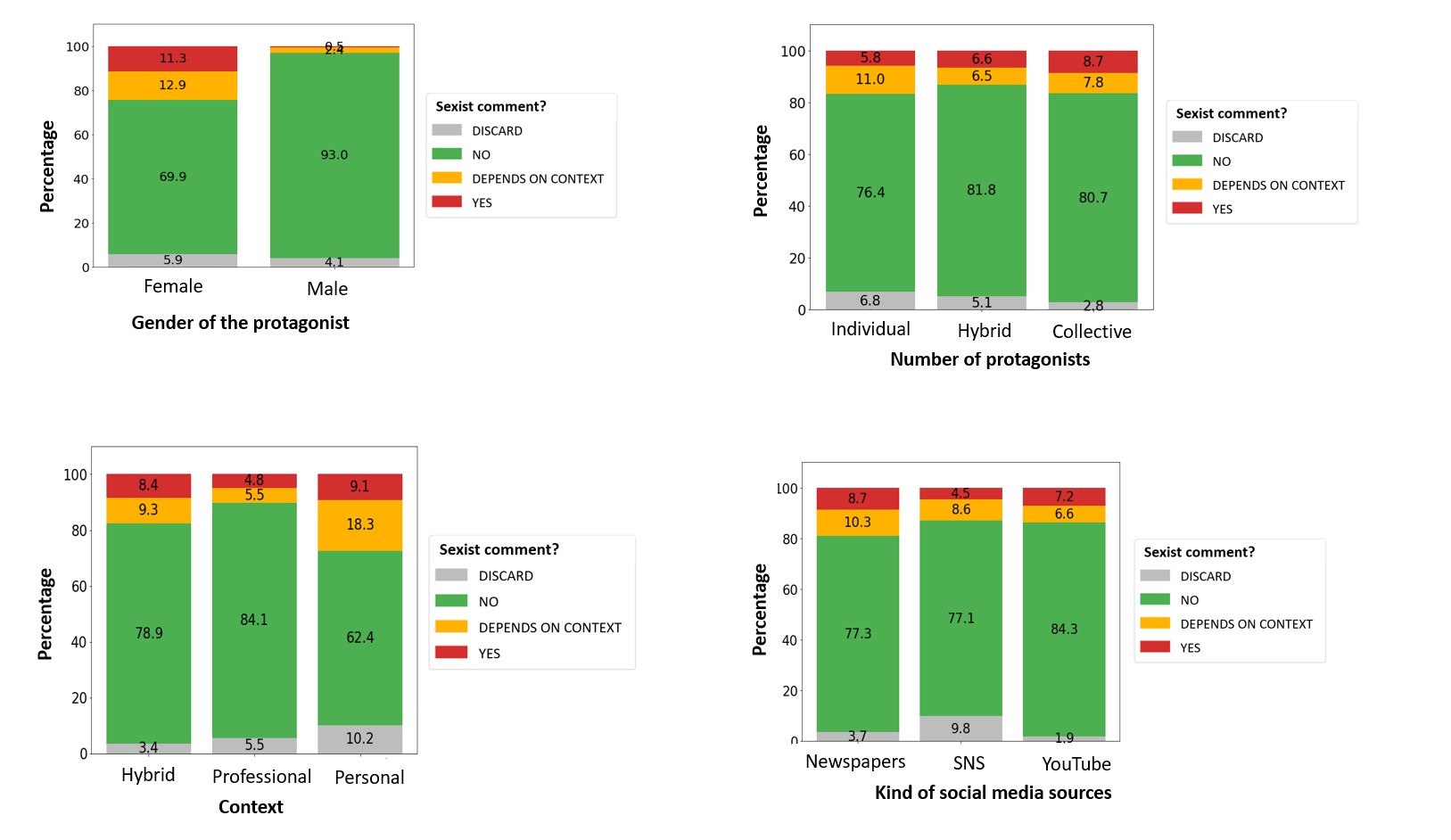}
\caption{{Characteristics}  of the training data~set.}
\label{fig:statistics2}
\end{figure}

After the labeling process, we obtained the results that are summarized in Figure~\ref{fig:statistics2}. In~the training set, a~clearly higher proportion of sexist comments are related to content where women are the main protagonist ($11.3\%$) and also when the content is related to a personal aspect ($9.1\%$). Sexist comments are also more frequent in newspapers and video comments, than~on Twitter, and also they are clearly higher when the content is related to a collective ($8.7\%$), rather than when it refers to an individual~woman.

\section{Phase 2: Model Definition and Training~Process}
\label{sec:model}

We decided to use an NLP/ML model to build the the \emph{anti-sexism alert} system, based on a supervised learning approach. Instead of creating an NLP/ML model from scratch, we decided to apply a transfer learning approach and use a pre-trained model, obtained from the Hugging Face model repository: the Hate-speech-CNERG/dehatebert-mono-spanish, which was pre-trained to detect hate speech in Spanish \cite{aluru2020deep}. This technique, using a pre-trained model, provides good results, and it is especially interesting in our case, because~we do not have a massive tagged data set (in Spanish). So it allowed us to obtain a good performance, since there are strong similarities between the language structures of hate speech and offensive sexist comments. Therefore, using this model as a base, we proceeded to perform a second training (or fine-tuning) of its parameters, using our training set of labeled~comments. 

The training process was performed using the manually labeled comments (4389). The~majority ({86.44\%}
) were manually labeled as ``Yes (sexist comment)'' or ``No (sexist comment)'', whereas $8.48$\% were labeled as ``Depends on the context'', and the last $5.08$\% was labeled as ``Discard''. The training process, performed using only the comments labeled as ``Yes/No'', was performed with a total of $3,794$ comments. They were split into two sub-sets: a training set (80\% of the tagged data) to train the NLP/ML model, and a test set (20\% of the tagged data), to evaluate the model. These percentages reveal a scenario of imbalance between labeled classes. To~deal with this issue, on~the one hand, we used an NLP model pre-trained on a context such as hate speech, which also naturally presents imbalanced classes. On~the other hand, when retraining and fitting this model with our database, we used an approach based on penalized models, so that class weights were taken into account.

Once the model was trained, we used the test data set to check the performance of the {\emph{anti-sexism alert}}  system. For~this evaluation, we used metrics described in Table~\ref{tab:performanceMetrics}, where $TP$ stands for true positives (sexist comments correctly classified), $FP$ means false positives (no sexist comments incorrectly classified), $TN$ stands for true negatives (no sexist comments correctly classified), and $FN$ means false negatives (sexist comments incorrectly classified). Consequently, accuracy represents the total percentage of cases that were correctly classified (identified) by the \emph{anti-sexism alert} system. Precision, measures the proportion of cases that, having been classified as positive, actually were positive. Recall, assesses the proportion of actually positive cases that were correctly classified, and $F_1 score$ corresponds to the harmonic mean of precision and recall.

 \begin{table}[htpb]
 \begin{center}
 \caption{{Performance} metrics.}
\begin{tabular}{lll} 
 Accuracy $= \frac{TP + TN}{TP + TNM + FP + FN}$  & Precision $= \frac{TP}{TP+FP}$ \\
 Recall $= \frac{TP}{TP+FN}$  & F\_1score $= 2 \cdot \frac{Precision \cdot Recall}{Precision + Recall}$ \\
\end{tabular}
\label{tab:performanceMetrics}
\end{center}
\end{table}

\section{Phase 3: Testing the Classification~System}
\label{sec:analysis}

\begin{figure}[htpb]
\centering
\includegraphics[scale = 0.50]{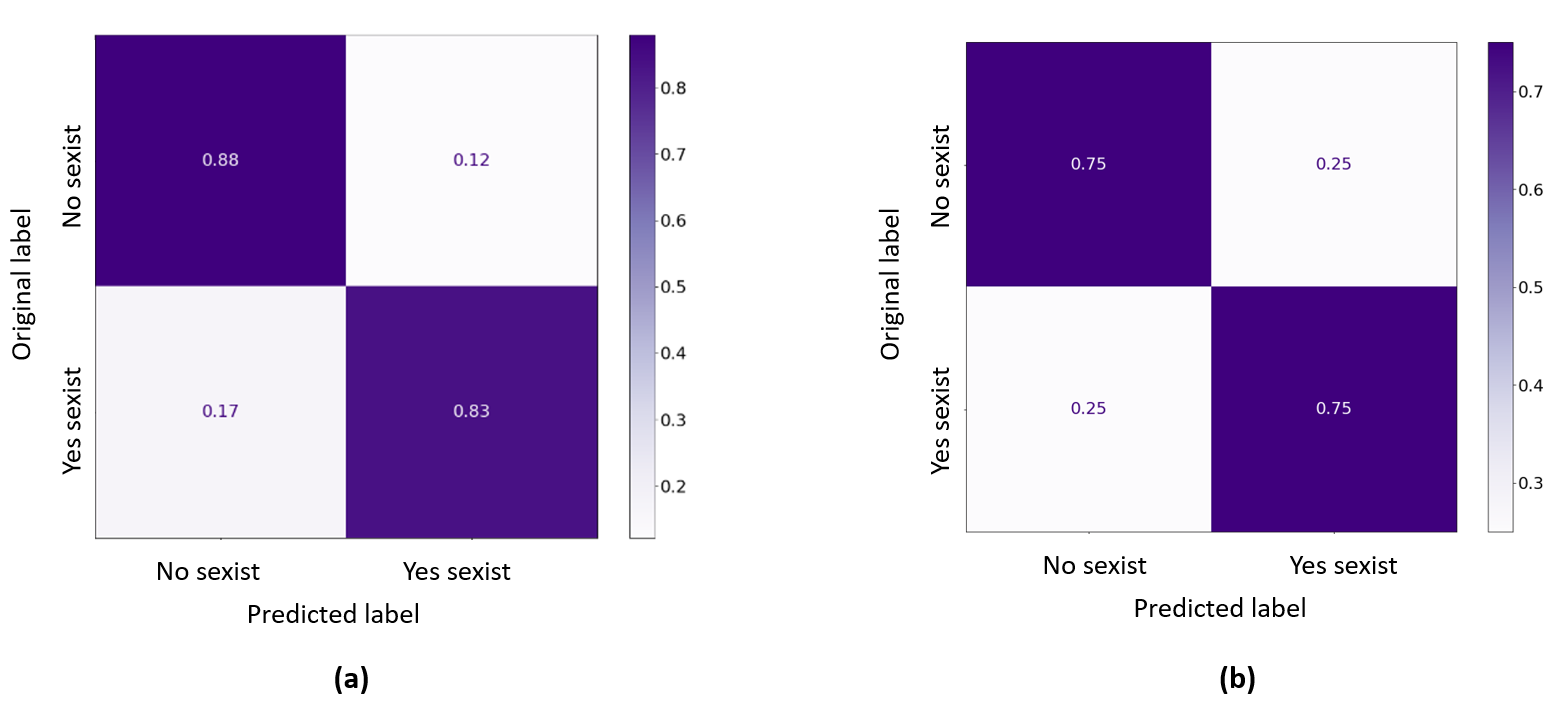}
\caption{Normalized confusion matrix (\textbf{a}) of the training set and (\textbf{b}) of the test~set.}
\label{fig:confusionmatrices}
\end{figure}

After training the model, we assessed the performance of the \emph{anti-sexism alert} system on the training data set and on the test data set. The obtained results are summarized in the precision, recall, and F1 scores of Tables~\ref{tab:trainingresults} and \ref{tab:testresults}, and the confusion matrices of Figure~\ref{fig:confusionmatrices}. These show the percentage of sexist/non-sexist comments correctly and incorrectly identified, in~(a) the training data set, and~(b) the test data set. As~expected, the~performance of the algorithm worsens when it has to classify the test set, composed of comments that have not been used in the training.

\begin{table}[htpb]
\begin{center}
\caption{{Training}  set: Results for sexist comments, non-sexist comments, and~both.\label{tab:trainingresults}}
\begin{tabular}{lccc} 
\textbf{Classification} & \textbf{Precision} & \textbf{Recall} & \textbf{F1\ score}  \\ 
\hline
No sexist & $0.84$ & $0.88$ & $0.86$  \\
Yes sexist & $0.87$ & $0.83$ & $0.85$  \\
Global & $0.86$ & $0.86$ & $0.86$ \\ 
\end{tabular}
\end{center}
\end{table}

\begin{table}[htpb]
\begin{center}
\caption{{Test}  set: Results for sexist comments, non-sexist comments, and~both.}
\begin{tabular}{lccc} 
\textbf{Classification} & \textbf{Precision} & \textbf{Recall} & \textbf{F1\ score} \\ 
\hline
No sexist & $0.75$ & $0.75$ & $0.75$  \\
Yes sexist & $0.75$ & $0.75$ & $0.75$  \\
Global & $0.75$ & $0.75$ & $0.75$  \\ 
\end{tabular}
\label{tab:testresults}
\end{center}
\end{table}

We were also interested in analyzing the sexist nature of the full set of comments related to each content. In~order to visualize the results in a simple and effective way, we defined a color-based \emph{anti-sexism alert} system that works as follows. First, it receives all the public comments linked to a content source (piece of news published in a newspaper, a~post on Twitter, or a video on YouTube). After~analyzing each one of the comments, it uses the typical colors of a traffic light to show the results: red to give an alert on sexist comments (if the proportion of sexist comments is higher than $5$\%), yellow to give an alert on potentially sexist comments (if the proportion of sexist comments is higher than $2.5$\% but lower than $5$\%), and green to highlight a non-sexist comment (if the proportion of sexist comments is lower than $2.5$\%). Of~course, these two thresholds ($5\%$ and $2.5\%$) may be modified. {Since the underlying objective is trying to avoid these comments becoming viral, these thresholds have to be carefully selected, in order to trigger an early alert.}

We used these thresholds to analyze the content sources of the test data set that have more than 100 comments, the~results are summarized in Table~\ref{tab:colorResults}: $13$ sources out of the original $59$ content sources. The~results of the NLP/ML classification system are summarized in Table~\ref{tab:colorResults}, where they are compared to the manual labeling. According to this, the color-based \emph{anti-sexism alert} system correctly classified $84.6\%$ of the comments of the analyzed content~sources. 

 Analyzing the results of applying the alert system on the test set, we found the following results. When a woman was the main protagonist, $28.6\%$ of the content sources were assigned as {\emph{green}} 
, $14.3\%$ were assigned as {\emph{yellow}}, 
 and $57.1\%$ were assigned as \emph{red}. However, when a man was the main protagonist, $100.00 \%$ of the content sources were declared as \emph{green}.

\begin{table}[htpb]
\begin{center}
\caption{{Results} 
 of color-based \emph{anti-sexism alert} system. {The content source ID is detailed in Table~\ref{tab:listSources}}.}
\begin{tabular}{l|cc|cc} 
& \multicolumn{2}{l}{\textbf{Manual Labeling}} & \multicolumn{2}{l}{\textbf{Prediction}} \\ 
\hline
\textbf{Content Source ID} & \textbf{\% Sexist} & \textbf{Color} & \textbf{\% Sexist} & \textbf{Color} \\ 
\hline
E17  & $0.00$ & { Green} & $0.77$ & { Green} \\
E2 & $2.97$ & { Yellow} & $1.98$ & { Green}\\
E3 & $1.06$ & { Green} & $2.66$ & { Green}\\
E5 & $43.38$ & { Red} & $41.18$ & { Red} \\ 
E7 & $0.00$ & { Green} & $1.36$ & { Green}\\
M1 & $8.59$ & { Red} & $7.03$ & { Red} \\
T10 & $6.73$ & { Red} & $4.81$ & { Yellow} \\
T13 & $0.00$ & { Green} & $0.00$ & { Green}\\
T17 & $1.94$ & { Green} & $0.97$ & { Green}\\
T19 & $0.00$ & { Green} & $0.00$ & { Green}\\
Y5 & $21.12$ & { Red} & $19.14$ & { Red} \\
Y7 & $0.00$ & { Green} & $1.77$ & { Green}\\ 
Y9 & $0.00$ & { Green} & $0.00$ & { Green}\\ 
\end{tabular}
\label{tab:colorResults}
\end{center}
\end{table}

\section{Discussion}
\label{sec:discussion}

After the different analyses performed (statistic analysis of the data set, statistic analysis of the labeled training data set, and the execution of the \emph{anti-sexism alert} system on the test data set), we obtained the following~conclusions.

First, we analyzed the distribution of the comments in the data set according to the different characteristics of the content sources. In~spite of having selected a similar number of content sources from the three types of social media (newspapers, Twitter, YouTube), the~number of comments is quite different among them (Figure \ref{fig:statistics1}): pieces of news from newspapers generated far fewer comments ($1.8\%$  of the total number of comments) than posts on Twitter ($80\%$) or videos on YouTube ($18.2\%$). 
This might be a consequence of the audience to which these three types of social media are addressed, with Twitter and YouTube being more interesting for youngsters, the~ones more keen on sharing their thoughts. Youtube users are younger than Twitter users (76\% of Youtube users are between 16 and 30 years old, compared to 45\% of Twitter users). Besides, only 8\% of Spanish internet users add comments on news items on news media websites. Participating on the news site was more common among men over 65 years old, whereas women are notably less participative than men, in all age ranges \cite{Espanha22}.

Analyzing the gender of the main protagonists, we have 30 content sources where men were protagonists vs. 29 contents sources that have women as the main protagonist (balanced sources). However, the~number of comments in the latter (54\%), were higher than in the former (46\%). Focusing on the number of the protagonists of the content sources, the~highest number of comments were obtained when the content source was about an individual person (54\%), whereas content sources about collectives generated less comments (14.7\%). Finally, content sources that mixed personal and professional aspects generated the highest number of comments ({45.5\%}
), while personal aspects generated less comments ({17.5\%} 
)

After that, we analyzed how the sexist comments were distributed in the training data set (manually labeled), according to the same characteristics (Figure \ref{fig:statistics2}). When the content source was about women (individual and/or collective) there was a relevant percentage of sexist comments ($11.3\%$) or comments that might be sexist ($12.9\%$). However, when the content source was about men (individual and/or collective) there is a residual percentage of sexist comments ($0.5\%$) or comments that might be sexist ($2.4\%$). These results reinforce other previous analyses, that stated women are mainly the object of sexist attacks on digital media \cite{charpak2015quality,molanes2018deep}. The~number of protagonists in the content source is not relevant for this study, since the percentage of sexist comments are quite balanced among the three categories: individual $5.8\%$, collective $8.7\%$, hybrid $6.6\%$. However, this result changes when we focus on the context of the content source: personal content obtains the highest percentage ($9.1\%$ of sexist comments and $18.3\%$ of potential sexist comments) versus professional content ($4.8\%$ of sexist comments and $5.5\%$ of potential sexist comments). Finally, the~type of source is also important: comments in the digital newspaper were more sexist ($8.7\%$ and $10.3\%$ of potentially sexists) than on Twitter ($4.5\%$ and $8.6\%$ of potentially sexists) or on YouTube ($7.2\%$ and $6.6\%$ of potentially sexists). This is especially remarkable, since the number of comments in the newspapers was the lowest (only $1.8\%$).

These differences could be related to the profile of the users of the media and networks used in the study. The~dominant profile of those who comment on news items in the digital press on the newspaper's own website is male and over 65 years old \cite{Espanha22}. In~contrast, the~average profile of social network users in Spain has an average age of around 40 and is equally likely to be male or female \cite{RRSS21}.

Finally, we analyzed the results obtained with the NLP/ML \emph{anti-sexism alert} system working on an individual basis, that is, analyzing each comment as stand-alone. According to the this analysis (Figure \ref{fig:confusionmatrices}), we can conclude the system had a very good performance on the training data set (in the specialized literature on NLP, it is considered that correct detection over $80\%$ is a very good result) and also a good performance on the test data~set. 

Additionally, the~color-based \emph{anti-sexism alert} system, which works on sets of comments (all the ones linked to a content source), also obtained a good result: the system was able to correctly classify $84.6\%$ of the analyzed content sources. What is more remarkable, the~mistakes were from yellow to green or vice~versa, but~it never happened that a green content source was assigned as red or vice~versa. 

Besides, we can also conclude that the gender of the main protagonist in the content source is key for the color-based \emph{anti-sexism alert} system: none of the content sources with men as the main protagonist provoked sexist comments, but~all the content sources considered as sexist were focused on women. This trend was also corroborated when the manual labeling was done. When the content source (pieces of news, video, or Tweet) was related to men, the sexist comments usually were posted as an answer to a comment made by a woman. However, when the content source was related to women, there were a higher number of comments about the protagonist that were considered to be sexist.


\section{Conclusions and Future~Work}
\label{sec:conclusions}

Digital sexism is increasingly worrying, because of the strong impact it has on the affected persons. There is no safe space when the aggression can be suffered at any time and any space, and additionally it can also become viral, amplifying the damage~caused.

Within this context, we propose an NLP/ML-based mechanism, to automatically identify sexist comments on social media. Our approach is able to (i) classify stand-alone comments, and also (ii) analyze the whole set of comments linked to a content source (piece of news, video, post, etc.). For~the latter, our algorithm works as a color-based \emph{anti-sexism alert} system, whose output is red, when the whole set of comments shows a clear sexist trend, green, when the proportion of sexist comments is insignificant, and~yellow~otherwise. 

We have defined an NLP/ML model based on a supervised learning approach, and we have applied transfer learning to avoid creating the solution from scratch. Thus, we used a pre-trained model to detect hate speech in Spanish (Hate-speech-CNERG/dehatebert-mono-spanish) and then we fine tuned it to re-train it with the specific characteristics of sexist comments. Since there is not adequate data sets in Spanish for this objective, we gathered a data set with comments from different source contents selected among three types of media sources: two newspapers, Twitter, and YouTube. After~that, we manually labeled part of the data set, in order to train the model and validate our~approach.

After performing the evaluation of our proposal, we can state that the results are promising, with good performance on the training set and on the test set.  As with any AI-based development, the~performance of the alert system is strongly conditioned by the size of the database used for training. For~this reason, we are currently working on an extension of the data set, including more comments and more sources.  Aligned with this purpose, we are developing an open website for the color-based \emph{anti-sexism alert} system, that would make possible to reach a wider audience and to perform further validation. Besides, we are working on creating user interfaces, aimed at using the \emph{anti-sexism alert} system as a tool for disseminating and raising awareness of sexist comments on social networks among young people. Finally, we are also working on defining a more complex scale (not only three values or colors), to give more precise information that could be used for a more ambitious objective: improving protocols to identify the real risk of gender-based violence by analyzing the messaging systems or social networks of a potential victim.  To sum up, artificial intelligence is often accused of replicating, and~even increasing, the~gender biases that exist in our society. However,~our results show that, a properly trained algorithm can detect violence in the language we use, and~flag it, preventing harm to victims. The~big companies that own social networks can certainly put ``AI for good'' into practice and try to put a stop to this violence, which hits women the hardest. It is in their power to do so, and~it is their responsibility.

\appendix

All the sources used to create the data set are listed in Table~\ref{tab:listSources}:
\begin{itemize}
    \item [-] Those identified as ``EX'', with $x$ being a number, belong to the digital newspaper {``El Mundo''} 
 (\url{https://www.elmundo.es/} (accessed on 20 January 2023)). 
    \item [-] Those identified as ``MX'', with $x$ being a number, belong to the digital sportive newspaper {``Marca''} 
 (\url{https://www.marca.com/} (accessed on 20 January 2023)). 
    \item [-] Those identified as ``TX'', with $x$ being a number, belong to the SNS {Twitter} 
 (\url{https://twitter.com/} (accessed on 20 January 2023)). 
    \item [-] Those identified as ``YX'', with $x$ being a number, belong to {YouTube} 
 (\url{https://www.youtube.com/} (accessed on 20 January 2023)). 
\end{itemize}

    \begin{table}[H]

    \caption{{List}  of content sources used to create the data~set.}\label{tab:listSources}
\begin{tabular}{ll} 

\textbf{ID} & \textbf{URL}  \\
E1 &  { \url{https://www.elmundo.es/loc/casa-real/2021/09/11/613c9e3efdddff381f8b4571.html} (accessed on 20 January 2023)} \\
E2 &  { \url{https://www.elmundo.es/economia/2021/09/10/613b40d3fc6c83b5728b466f.html} (accessed on 20 January 2023)} \\
E3 & { \url{https://www.elmundo.es/cultura/cine/2021/09/11/613cf13921efa039798b45c2.html} (accessed on 20 January 2023)} \\ 
E4 & { \url{https://www.elmundo.es/metropoli/teatro/2021/09/04/61321f1dfdddff02358b46a0.html}} \\ 
E5 & { \url{https://www.elmundo.es/espana/2021/08/30/612cb74dfc6c8398258b4620.html} (accessed on 20 January 2023)} \\ 
E6 & { \url{https://www.elmundo.es/andalucia/2021/10/14/6167febafdddff2c728b456e.html} (accessed on 20 January 2023)} \\ 
E7 & { \url{https://www.elmundo.es/economia/vivienda/2021/10/14/61680e82fc6c8317098b45dc.html} (accessed on 20 January 2023)} \\
E8 & { \url{https://www.elmundo.es/television/momentvs/2021/10/12/61651442e4d4d83f698b45ba.html} (accessed on 20 January 2023)} \\ 

E9 & { \url{https://www.elmundo.es/television/momentvs/2021/10/05/615be7d8e4d4d84c168b456f.html} (accessed on 20 January 2023)} \\ 
E10 & { \url{https://www.elmundo.es/deportes/baloncesto/2021/10/06/615cab93fc6c83516f8b45c1.html} (accessed on 20 January 2023)} \\ 
E11 & { \url{https://www.elmundo.es/loc/famosos/2021/07/06/60e2eb66e4d4d83c128b4636.html} (accessed on 20 January 2023)} \\ 
E12 & { \url{https://www.elmundo.es/loc/famosos/2020/11/25/5fbd6003fdddff43a18b460c.html} (accessed on 20 January 2023)} \\ 
E13 & { \url{https://www.elmundo.es/loc/famosos/2021/04/27/6087c60cfc6c832c358b45aa.html} (accessed on 20 January 2023)} \\ 
E14 & { \url{https://www.elmundo.es/economia/2021/09/08/6138960121efa0855a8b45b2.html} (accessed on 20 January 2023)} \\ 
E15 & { \url{https://www.elmundo.es/cultura/cine/2021/07/19/60f560f8e4d4d8297b8b458f.html} (accessed on 20 January 2023)} \\ 
E16 & { \url{https://www.elmundo.es/metropoli/teatro/2021/09/03/61264d4ae4d4d8284c8b45a5.html} (accessed on 20 January 2023)} \\ 
E17 & { \url{https://www.elmundo.es/espana/2021/11/08/6187c214fc6c83cc328b4582.html} (accessed on 20 January 2023)} \\ 
E18 & { \url{https://www.elmundo.es/andalucia/2021/10/18/616d4082fc6c834d028b45a6.html?intcmp=MNOT23801&s_kw=3} (accessed on \linebreak 20 January 2023)} \\ 
E19 & { \url{https://www.elmundo.es/television/2021/08/16/611a0fadfdddff45608b45a2.html} (accessed on 20 January 2023)} \\ 
E20 & { \url{https://www.elmundo.es/f5/descubre/2021/08/12/6115045afc6c8360698b45e2.html} (accessed on 20 January 2023)} \\ 
E21 & { \url{https://www.elmundo.es/television/2021/08/06/610cdad5fc6c83a4478b45e5.html} (accessed on 20 January 2023)} \\ 
E22 & { \url{https://www.elmundo.es/loc/famosos/2021/10/07/615edbf8e4d4d8b7688b4577.html} (accessed on 20 January 2023)} \\

M1 & { \url{https://www.marca.com/tiramillas/television/2021/09/07/6137c0f3ca4741ff738b45e6.html} (accessed on 20 January 2023)} \\
M2 & { \url{https://www.marca.com/tenis/open-australia/2021/02/08/6020bf13268e3e964f8b459c.html} (accessed on 20 January 2023)} \\
M3 & { \url{https://www.marca.com/tiramillas/musica/2021/10/29/617bd33646163f20578b45fc.html} (accessed on 20 January 2023)} \\
M4 & { \url{https://www.marca.com/futbol/futbol-femenino/2021/11/02/61816bebe2704e4ea58b459e.html} (accessed on 20 January 2023)} \\
M5 & { \url{https://www.marca.com/tiramillas/television/2021/09/02/6130727122601da10a8b4600.html?intcmp=MNOT23801&s_kw=2} (accessed on 20 January 2023)} \\
M6 & { \url{https://www.marca.com/tiramillas/musica/2021/10/19/616e9c5a22601d1c078b45da.html?intcmp=MNOT23801&s_kw=1} (accessed on \linebreak 20 January 2023)} \\
M7 & { \url{https://www.marca.com/tiramillas/musica/2021/10/02/61588311268e3e851b8b45c1.html} (accessed on 20 January 2023)} \\

\end{tabular}
\end{table}

\begin{table}[H]
\caption{{List}  of content sources used to create the data~set.}\label{tab:listSources2}
\begin{tabular}{ll} 
\textbf{ID} & \textbf{URL}  \\

T1 & {\url{https://twitter.com/search?q=\%23ballenadevallecas\&src=typed_query\&f=top} (accessed on 20 January 2023)} \\
T2 & {\url{https://twitter.com/search?q=\%23pedrochecampanadas\&src=typed_query\&f=top} (accessed on 20 January 2023)} \\
T3 & {\url{https://twitter.com/search?q=\%23chicotecampanadas\&src=typed_query\&f=top} (accessed on 20 January 2023)} \\
T4 & {\url{https://twitter.com/search?q=(\%23manuelacarmena)\%20OR\%20(\%40ManuelaCarmena)\&src=typed_query\&f=top} (accessed on \linebreak 20 January 2023)} \\
T5 & {\url{https://twitter.com/search?q=(\%23anapastor)\%20OR\%20(\%40_anapastor_)\&src=typed_query&f=top} (accessed on 20 January 2023)} \\
T6 & {\url{https://twitter.com/search?q=(\%23carlotacorredera)\%20OR\%20(\%40CarlotaLlauger)\&src=typed_query&f=top} (accessed on \linebreak 20 January 2023)} \\
T7 & {\url{https://twitter.com/search?q=(\%23rociocarrasco)\%20OR\%20(\%40RocioCarrascooo)\&src=typed_query&f=top} (accessed on 20 January 2023)} \\
T8 & {\url{https://twitter.com/search?q=(\%23antoniodavid)\%20OR\%20(\%40adavidflores)\&src=typed_query&f=top} (accessed on 20 January 2023)} \\
T9 & {\url{https://twitter.com/search?q=\%23rocioyonotecreo\&src=typed_query\&f=top} (accessed on 20 January 2023)} \\
T10 & {\url{https://twitter.com/search?q=(\%23irenemontero)\%20OR\%20(\%40IreneMontero)\&src=typed_query\&f=top} (accessed on 20 January 2023)} \\
T11 & {\url{https://twitter.com/search?q=(\%23juanarivas)\%20OR\%20(\%40JuanaRivasLucha)\&src=typed_query&f=top} (accessed on 20 January 2023)} \\
T12 & {\url{https://twitter.com/stopfeminazis00} (accessed on 20 January 2023)} \\
T13 & {\url{https://twitter.com/search?q=(\%23albertochicote)\%20OR\%20(\%40albertochicote)\&src=typed_query\&f=top} (accessed on 20 January 2023)} \\
T14 & {\url{https://twitter.com/search?q=antonioferreras&src=typed_query} (accessed on 20 January 2023)} \\
T15 & {\url{https://twitter.com/search?q=(\%23almeida)\%20OR\%20(\%40AlmeidaPP_)\&src=typed_query\&f=top} (accessed on 20 January 2023)} \\
T16 & {\url{https://twitter.com/search?q=(\%23jorgejavier)\%20OR\%20(\%23jorgejaviervazquez)\%20OR\%20(\%40jjaviervazquez)\&src=typed_query\&f=top} (accessed on 20 January 2023)} \\
T17 & {\url{https://twitter.com/search?q=(\%23pedrosanchez)\%20OR\%20(\%40sanchezcastejon)\&src=typed_query\&f=top} (accessed on 20 January 2023)} \\
T18 & {\url{https://twitter.com/search?q=(\%23landroberTVG)\%20OR\%20(\%40landroberTVG)\&src=typed_query\&f=top} (accessed on 20 January 2023)} \\
T19 & {\url{https://twitter.com/search?q=(\%23pablocasado)\%20OR\%20(\%40pablocasado_)\&src=typed_query\&f=top} (accessed on 20 January 2023)} \\
T20 & {\url{https://twitter.com/hashtag/lamareaazulnosecansa?src=hashtag_click} (accessed on 20 January 2023)} \\

Y1 & {\url{https://www.youtube.com/watch?v=JJhidQUN3oI} (accessed on 20 January 2023)}  \\
Y2 & {\url{https://www.youtube.com/watch?v=EviNPfSmoQs} (accessed on 20 January 2023)}  \\
Y3 & {\url{https://www.youtube.com/watch?v=lVADjbxNRFA} (accessed on 20 January 2023)}  \\
Y4 & {\url{https://www.youtube.com/watch?v=xEi4WJVdzBM} (accessed on 20 January 2023)}  \\
Y5 & {\url{https://www.youtube.com/watch?v=aQu81Wsar6Y} (accessed on 20 January 2023)}  \\
Y6 & {\url{https://www.youtube.com/watch?v=cgeDqdMDYrQ&ab_channel=La2} (accessed on 20 January 2023)}  \\
Y7 & {\url{https://www.youtube.com/watch?v=GYwje9tSrro&ab_channel=JoanPlanas} (accessed on 20 January 2023)} \\ 
Y8 & {\url{https://www.youtube.com/watch?v=UxuKc_Oe_k0&ab_channel=RomaGallardo} (accessed on 20 January 2023)}  \\
Y9 & {\url{https://www.youtube.com/watch?v=gA8hUBDCPkk&ab_channel=laSexta} (accessed on 20 January 2023)} \\ 
Y10 & {\url{https://www.youtube.com/watch?v=R8TEBqy8-As} (accessed on 20 January 2023)}  \\

\end{tabular}
\end{table}

\end{document}